\documentclass[superscriptaddress,showpacs,prb,12pt,endfloats]{revtex4-1}
\usepackage[utf8]{inputenc}
\usepackage{amsmath,amssymb}
\usepackage{graphicx,textcomp}
\usepackage{color}
\usepackage{todonotes}
\usepackage{todonotes}
\usepackage[utf8]{inputenc}
\usepackage{braket}
\usepackage[normalem]{ulem}

\setcitestyle{numbers,square}

\definecolor{ftcol}{rgb}{0.75,0.25,0.0}

\begin{document}
\title{Thermodynamic and electronic properties of ReN$_2$ polymorphs at high-pressure}

\author{Ferenc Tasnádi}
\affiliation{Department of Physics, Chemistry, and Biology (IFM),
Link\"{o}ping University, SE-581 83, Link\"oping, Sweden.}

\author{Florian Bock}
\affiliation{Department of Physics, Chemistry, and Biology (IFM),
Link\"{o}ping University, SE-581 83, Link\"oping, Sweden.}

\author{Alena Ponomareva}
\affiliation{Materials Modeling and Development Laboratory, NUST ``MISIS'', 119049 Moscow, Russia.}

\author{Maxim Bykov}
\affiliation{Bayerisches Geoinstitut, University of Bayreuth, 95440 Bayreuth, Germany}

\author{Saiana Khandarkhaeva}
\affiliation{Bayerisches Geoinstitut, University of Bayreuth, 95440 Bayreuth, Germany}
\affiliation{Material Physics and Technology at Extreme Conditions, Laboratory of Crystallography, University of Bayreuth, Universit\"atstrasse 30, 95440 Bayreuth, Germany.}

\author{Leonid Dubrovinsky}
\affiliation{Bayerisches Geoinstitut, University of Bayreuth, 95440 Bayreuth, Germany}

\author{Igor A. Abrikosov}
\affiliation{Department of Physics, Chemistry, and Biology (IFM),
Link\"{o}ping University, SE-581 83, Link\"oping, Sweden.}
\affiliation{Materials Modeling and Development Laboratory, NUST ``MISIS'', 119049 Moscow, Russia.}

\date{\today}

\begin{abstract}
High pressure synthesis of rhenium nitride pernitride ReN$_2$ with crystal structure
unusual for transition metal dinitrides and high values of hardness and bulk modulus attracted significant
attention to this system.
We investigate the thermodynamic and electronic properties of the P2$_1$/c
phase of ReN$_2$ and compare them with two other polytypes, C2/m and P4/mbm phases
suggested in the literature. Our calculations of the formation enthalpy at zero temperature
show that the former phase is the most stable of the three up to pressure p=170\,GPa,
followed by the stabilization of the P4/mbm phase at higher pressure. 
The theoretical prediction is confirmed by diamond anvil cell synthesis of 
the P4/mbm ReN$_2$ at $\approx$175\,GPa.
Considering the effects of finite temperature in the quasi-harmonic approximation at p=100 GPa
we demonstrate that
the P2$_1$/c phase has the lowest free energy of formation at least up to 1000\,K.
Our analysis of the pressure dependence of the electronic structure of the rhenium nitride pernitride
shows a presence of two electronic topological transitions around 18\,GPa, when 
the Fermi surface changes its topology due to an appearance of a new
electron pocket at the high-symmetry Y$_2$ point of the Brillouin zone
while the disruption of a neck takes place slightly off from the $\Gamma$-A line.
\end{abstract}
\maketitle
\section{Introduction} 
High pressure diamond anvil cell (DAC) experiment is a successful approach to establish wide variety
of physical conditions for synthesizing materials
\cite{mcmillan_new_2002,bykov_fe-n_2018,dubrovinskaia_terapascal_2016}. Though, exploring metastable
phases of materials, which is a challenging experimental task because one has to achieve
control of the small free energy barriers separating different polymorphs.
On the other hand, computational high-throughput approaches \cite{ludwig_discovery_2019,xia_novel_2018} with
sophisticated structure prediction algorithms \cite{glass_uspexevolutionary_2006,pickard_ab_2011} have
entered into the field to advance the materials discoveries.
The launch of the materials genome initiative (MGI) \cite{de_pablo_new_2019} has further accelerated this
trend \cite{agrawal_perspective_2016,sun_thermodynamic_2017} and triggered
the need to develop infrastructure to store and utilize the data calculated for different
materials. Large databases, such as  NOMAD \cite{draxl_nomad_2019}, Materials Project \cite{jain_commentary_2013},
AFLOW \cite{curtarolo_aflow_2012}, BioExcel \cite{andrio_bioexcel_2019},
Topological Material DataBase \cite{vergniory_complete_2019}, etc. have been created, which contain different properties of both,
existing and hypothetical materials.
Analyzing the data from Materials Project database Sun et al. observed that nitrides have the largest thermodynamic
scale of metastability defined as the energy differences between stable and metastable structures, $\approx$190 meV/atom
\cite{sun_thermodynamic_2016}, making this class of materials very interesting for further
experimental and theoretical exploration. 

At the same time, DAC has been actively used to synthesize
novel nitrogen rich phases of transition metal nitrides \cite{bykov_fe-n_2018,bykov_high-pressure_2018}.
One applies high pressure and temperature
in the chamber filled with molecular nitrogen or azides (AN$_3$, A=Li, Na, etc.) and metals, such as
rhenium, tungsten, osmium etc. \cite{bykov_high-pressure_2020,bykov_high-pressure_2019}. The synthesized metastable and nitrogen rich materials -
polymeric forms of nitrogen chains have also been observed \cite{bykov_high-pressure_2018}, represent a class of high-energy-density
materials often with superior mechanical properties \cite{bykov_high-pressure_2019}.

One challenge of using the high-pressure  experiment to discover novel materials with attractive properties
is to quench them to ambient conditions. The work by Bykov et al. \cite{bykov_high-pressure_2019} have shown that ReN$_2$ compound in
the monoclinic P2$_1$/c phase discovered in a DAC experiment can be also synthesized in larger amount in a
large-volume press at lower pressure. 
This compound has a crystal structure unusual for transition metal dinitrides MN$_2$. It contains covalently bound
dinitrogen dumbbells and discrete nitrogen atoms and represents an example of a mixed nitride-pernitride compound.
Quenched to ambient pressure, the rhenium nitride pernitride showed
high mechanical properties, hardness of 36.7(8) GPa and very high value of the bulk moduli of 428(10)GPa.
The P2$_1$/c phase was not reported in earlier experiments. Remarkably, despite numerous theoretical studies of
the Re-N system at this composition, it was not predicted theoretically even with the use of advanced structure prediction
algorithms.

Kawamura et al. \cite{kawamura_synthesis_2012} has reported the synthesis of ReN$_2$ at 7.7\,GPa and 1473-1873\,K, with
hexagonal P6$_3$/mmc (MoS$_2$ type) structure. Elastic and mechanical properties of the phase have
been investigated using T=0\,K density functional theory (DFT) calculations \cite{shein_structural_2013}.
In the same year, Du et al. \cite{du_investigation_2010} based on DFT calculations has proposed the
tetragonal P4/mmm phase of ReN$_2$ underlined by the existence of the same phase for ReB$_2$
\cite{chung_synthesis_2007}. However, static 0\,K DFT calculations by Wang et al.
\cite{wang_does_2013} have ruled out the P4/mmm phase as ground state of ReN$_2$.  Instead, the calculations
have indicated that the monoclinic C2/m structure of ReN$_2$ is more stable at 0\,K than the experimentally
found P6$_3$/mmc phase. Furthermore, it has been shown that above 130\,GPa the P4/mbm phase becomes the
favoured phase. A computational structural search for stable and metastable rhenium-nitrides
up to 100 GPa pressures has been conducted \cite{zhao_nitrogen_2015} using a sophisticated evolutionary algorithm implemented
into USPEX \cite{glass_uspexevolutionary_2006}. The study has confirmed the C2/m phase of ReN$_2$ as ground state between 10 and
100 GPa. The P4/mmm, Pbcn and P6$_3$/mmc phases have been found metastable in the investigated pressure
range.

In the present paper we investigate the thermodynamic stability of the P2$_1$/c phase
with respect to the competing tetragonal P4/mbm and C2/m monoclinic phases in the pressure range between 0 GPa and 180 GPa.
We use first principles electronic structure calculations and a quasi-harmonic approximation for the lattice dynamics and
establish that the former is indeed thermodynamically more stable then two other polymorphs at pressure up to $\approx$170 GPa.
At higher pressure the calculations predict the stabilization of the P4/mbm phase.
As this phase was not reported in earlier experiments, we carry out the high-pressure synthesis of ReN$_2$ in diamond anvil
cell at $\approx$175\,GPa.
The theoretical prediction is verified by a characterization ot the synthesized sample, which confirms the stabilization of
the P4/mbm ReN$_2$. 
Moreover, we calculate electronic properties of the P2$_1$/c phase and show the presence of two electronic topological transitions
at $\approx$ 18\,GPa.
\section{Computational details}
Simulations of the phase stability at T=0\ K have been performed using the Quantum Espresso
(QE) program package \cite{giannozzi_quantum_2009} with PAW pseudo-potentials \cite{blochl_projector_1994} using 60 Rydberg for kinetic energy
cutoff and 350 Rydberg for the density and potential cutoff. The exchange correlation energy
was approximated by the Perdew-Burke-Ernzerhof generalized gradient functional (PBE-GGA) \cite{perdew_generalized_1996}. In case
of the P4/mbm structure we have used a $(16\times 16\times 23)$ k-mesh for sampling the Brillouin zone and
the equation of state has been derived through fitting of the calculated total energies at different volumes by the third
order Burch-Murnighan expression. The two monoclinic structures
(C2/m and P2$_1$/c) have been relaxed by applying the variable
cell shape method introduced by Wentzcovitch \cite{wentzcovitch_ab_1993} using a ($18\times 10\times 14$) sampling of the Brillouin
zone. Density of states has been calculated after doubling the k-mesh density. Fermi surfaces have been calculated and visualized
using Xcrysden \cite{kokalj_xcrysdennew_1999} and VESTA \cite{momma_vesta_2011}.

The free energies of formation have been calculated within the quasi-harmonic approximation
using Phonopy \cite{togo_first_2015} combined with force calculations performed with VASP
\cite{kresse_ab_1993,kresse_ab_1994,kresse_efficiency_1996,kresse_efficient_1996}. 
In all VASP calculations the energy cutoff was set to 700\,eV.
Table \ref{tab_1} lists the supercell sizes and the applied sampling of the Brillouin zone in the VASP calculations.
To underline the agreement of our simulations by QE and VASP,
we have compared the stress tensor elements of the three different phases at around 100 GPa with unrelaxed atomic positions in the unit cells
using the supercell sizes and k point samplings from Table\,\ref{tab_1}.
The deviation are less than 4 GPa. The largest error, 4\%, has been observed for $\sigma_{yy}$ in case of comparing the
P2$_1$c and C2/m phases. In the enthalpy calculations the hcp phase of rhenium \cite{anzellini_equation_2014} and the cubic gauche
phase of nitrogen (which is stable up to $\approx$150 GPa) have been used as the end-member states \cite{laniel_high-pressure_2020}.

\section{Experiments}
The piece of Re metal was placed in a sample chamber of a BX90 diamond anvil cell \cite{kantor_bx90_2012} equipped with Boehler-Almax
type diamonds (40 $\mu$m culet diameters).
Nitrogen served as a pressure-transmitting medium and as a reagent \cite{kurnosov_novel_2008}. The DAC was compressed up to the target
pressure of $\approx$ 175\,GPa and laser-heated using double sided laser-heating systems installed at the at the Bayerisches
Geoinstitute (BGI, Bayreuth, Germany) \cite{fedotenko_laser_2019}.
The sample was studied by means of powder and single-crystal X-ray diffraction at the synchrotron beamline ID11 of the ESRF using
a monochromatic X-ray beam focused to $\approx 0.3 \times 0.3$ $\mu$m$^2$. For the single-crystal XRD measurements the sample was
rotated around the vertical $\omega$-axis in a range $\pm$38$^{\circ}$. The diffraction images were collected with an angular step
$\Delta\omega$ = 0.5$^{\circ}$ and an exposure time of 10\,s/frame. For analysis of the single-crystal diffraction data
(indexing, data integration, frame scaling and absorption correction) we used the CrysAlisPro software package. To calibrate an
instrumental model in the CrysAlisPro software, i.e., the sample-to-detector distance, detector’s origin, offsets of goniometer angles,
and rotation of both X-ray beam and the detector around the instrument axis, we used a single crystal of orthoenstatite
(Mg$_{1.93}$Fe$_{0.06}$)(Si$_{1.93}$,Al$_{0.06}$)O$_6$, Pbca space group, a = 8.8117(2), b = 5.18320(10), and c = 18.2391(3)\,\AA.
The structure was solved based on the single-crystal XRD data using the ShelXT structure solution program \cite{sheldrick_shelxt_2015} by intrinsic phasing and
refined using the Olex2 program \cite{dolomanov_olex2_2009}. Pressure was determined based on the equation of state of Re. The refined unit cell volume of Re
22.50(4) \AA\ may correspond to pressures of 172.5\,GPa (Anzellini et al. \cite{anzellini_equation_2014}), 177\,GPa
(Zha et al. \cite{zha_rhenium_2004}) or even 206\,GPa (Dubrovinsky et al. \cite{dubrovinsky_implementation_2012}).

\begin{table}[h]
\centering
\begin{tabular}{c|c|c|c|c}
structure & supercell & kpoint-mesh& supercell& kpoint-mesh\\
\hline
cg-N & $(1\times 1\times 1)$ & $(23\times 23\times 23)$ & $(4\times 4\times 4)$ & $(5\times 5\times 5)$\\
hcp Re & $(1\times 1\times 1)$ & $(15\times 15\times 10)$ & $(5\times 5\times 4)$ & $(5\times 5\times 5)$\\
P4/mbm ReN$_2$ & $(1\times 1\times 1)$ & $(19\times 19\times 27)$ & $(3\times 3\times 5)$ & $(5\times 5\times 5)$\\
C2/m ReN$_2$ & $(1\times 1\times 1)$ & $(9\times 21\times 7)$ & $(4\times 4\times 2)$ & $(5\times 5\times 5)$\\
P2$_1$/c ReN$_2$ & $(1\times 1\times 1)$ & $(24\times 14\times 18)$ & $(3\times 2\times 3)$ & $(3\times 3\times 3)$\\
\end{tabular}
\caption{The size of the structural models and the appied sampling of the Brillouin zone to calculate formation
enthalpy and the phonon dispersions.}
\label{tab_1}
\end{table}

\section{Results}

\begin{figure}
\includegraphics[width=8cm]{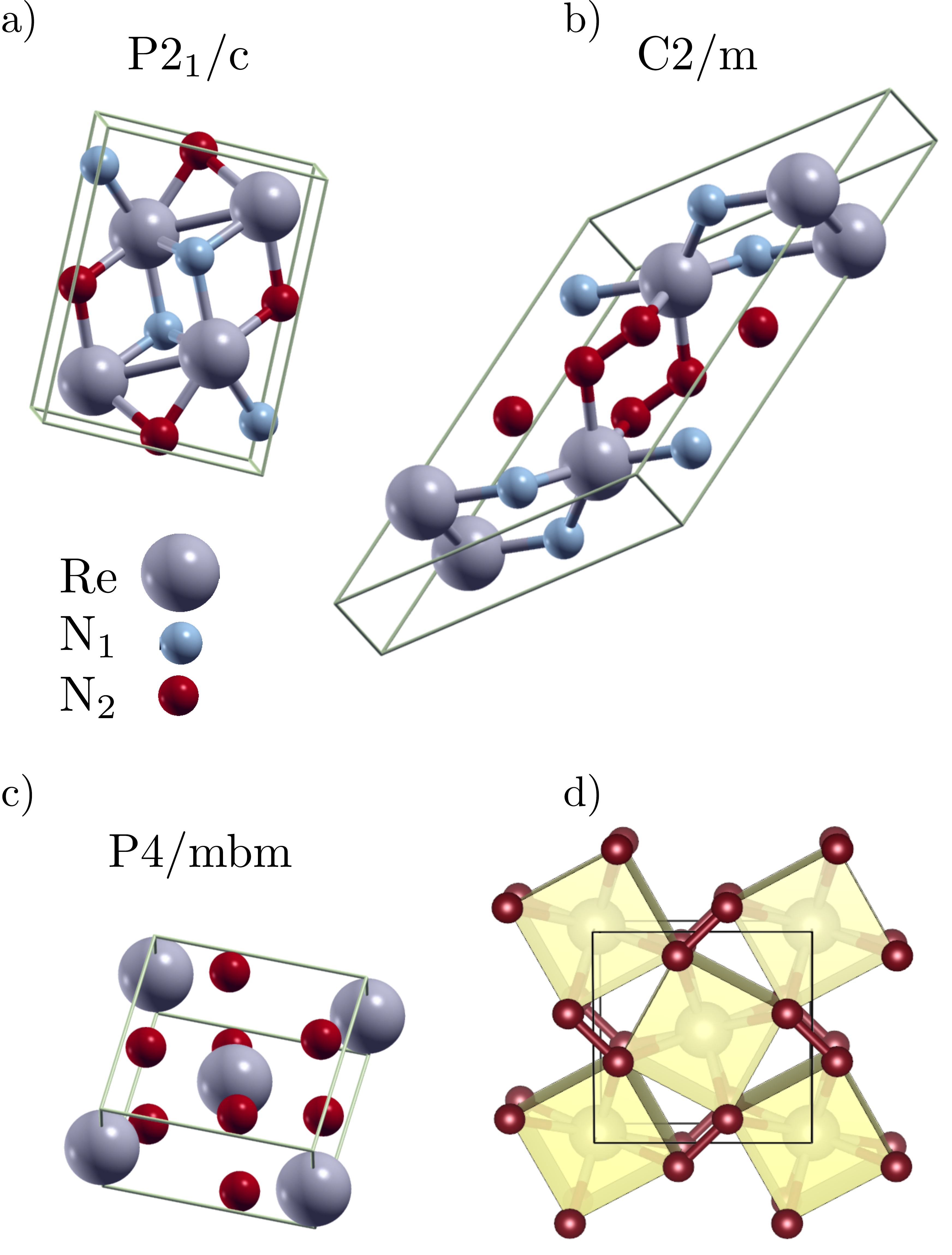}
\caption{The P2$_1$/c, C2/m and P4/mbm phases of ReN$_2$. Big circles show the Re atoms, while the small
ones the nitrogens singles (N$_1$) and dumbells (N$_2$). Panel d) shows the structure of the P4/mbm structure
and the formed ReN$_8$ rectangular prisms at $\approx$175 GPa as seen along the $c$-axis.}
\label{fig_1}
\end{figure}

Figure \ref{fig_1} shows three polymorphs of ReN$_2$ considered in this study: P2$_1$/c, C2/m and P4/mbm.
The large sized spheres represent the rhenium atoms, while the small ones the nitrogen atoms or dumbbells.
The discrete nitrogen atoms are denoted
by N1. If a nitrogen atom binds to another one and they form a dumbbell then both nitrogen atoms are labeled by N$2$.
The P4/mbm structure contains only nitrogen dumbbells, while the two monoclinic structures have both types of
nitrogens. Table \ref{tab_2} summarizes the optimized structural parameters obtained at T=0\,K and $p=0$\,GPa.

\begin{table}[h]
\centering
\begin{tabular}{c|c|l}
\text{structure}&space group&lattice parameters (\AA)\\
\hline
\text{P2$_{1}$/c}&14 (moniclinic)&a=3.630\ ,b=6.432,\ c=4.977,\ $\beta$=111.511$^{\circ}$\\
\text{C2$_{\text{m}}$} &12 (moniclinic)&a=6.800,\ b=2.824,\ c=9.336,\
$\beta$=142.35$^{\circ}$\\
\text{P4/mbm}&127 (tetragonal)&a=b=4.376,\ c=2.650\\
\end{tabular}
\caption{Space group numbers and the lattice parameters of the investigated
ReN$_2$ phases at T=0\,K and p=0 GPa.}
\label{tab_2}
\end{table}

\begin{figure}
\includegraphics[width=9cm]{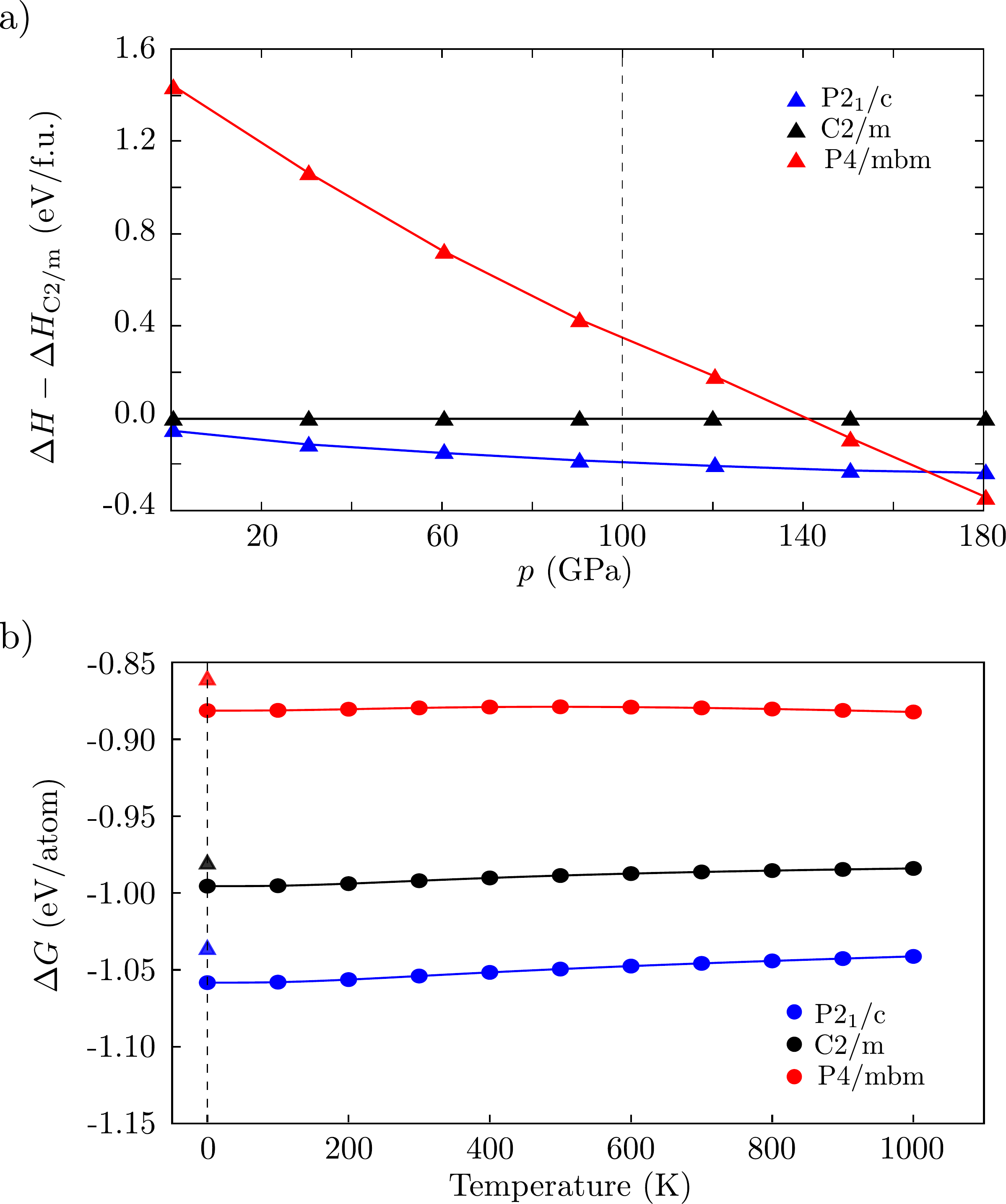}
\caption{a) Formation entalpy differences of the P2$_1$/c, C2/m and P4/mbm phases of ReN$_2$ wrt. the C2/m phase.
b) Gibbs free energy of formation of the P2$_1$/c, C2/m and P4/mbm phases at 100\,GPa within the quasi-harmonic
approximation up to 1000\,K. The dashed vertical line is
used to make a correspondence between a) and b) - see the text.}
\label{fig_2}
\end{figure}

In Figure \ref{fig_2}a we present the calculated formation enthalpy differences ($\Delta H$) 
for the P2$_1$/c and P4/mbm phases in the pressure interval from 0 GPa
to 180 GPa relative to the values obtained for the C2/m  phase.
The relevance of spin-orbit coupling (SOC) effects on the formation energy differences between the
P2$_1$/c and C2/m phases at 0 GPa can be ruled out by the electronic band structures calculated without
SOC, see Materials Project \cite{jain_commentary_2013} IDs:  mp-1077354 and mp-1019055.
The figure shows that the recently sythesized P2$_1$/c phase is the the most stable of the three up to
$\approx$170\,GPa at T=\,0 K. Above this pressure the tetragonal phase is favored over the two monoclinic
structures. Another interesting observation is that the P2$_1$/c phase is favored over the C2/m
monoclinic phase in the whole pressure range. The dynamical stability at 0 GPa has been proven for
all the three phases earlier \cite{yan_theoretical_2013,zhao_nitrogen_2015,bykov_high-pressure_2019}.
The effect of temperature on the relative stability of the three phases is analyzed in Fig.\ref{fig_2}b).
The figure shows the Gibbs free energy of formation of
the P2$_1$/c, C2/m and P4/mbm phases at 100\,GPa up to 1000\,K. The dashed vertical line in Fig.\ref{fig_2}a) and b) is
used to establish the correspondence between the zero temperature static calculations (triangles) and calculations
including the effects of the lattice dynamics, e.g. the zero-point motion (circles).
One sees that the relative order of the three phases is not changed by temperature, though the
formation energy differences become slightly smaller with increasing temperature. In Fig. \ref{fig_2}b) along the
dashed line the energy differences between the values marked by a triangle and a sphere shows the effect of the
zero-point energies of the lattice vibrations for each of the phases.

\begin{figure}
\includegraphics[width=16cm]{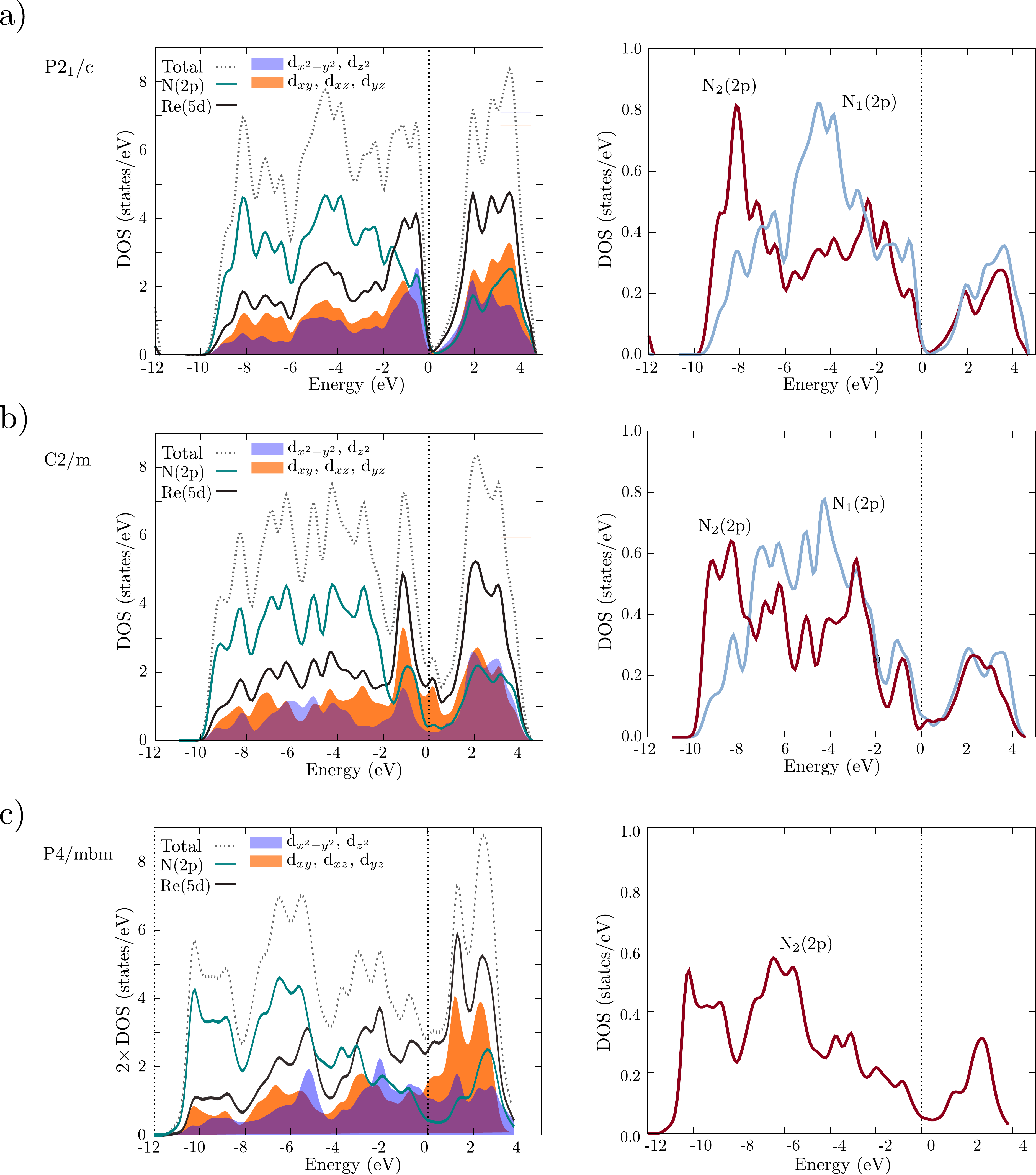}
\caption{Calculated total and partial density of states in
P2$_1$/c a), C2/m b) and P4/mbm c) phases of ReN$_2$ at 0 GPa and T=0\,K.}
\label{fig_3}
\end{figure}

\begin{table}[h]
\centering
\begin{tabular}{l|l}
\hline
\multicolumn{2}{l}{\bf Crystal data}\\
Chemical formula & ReN$_2$ \\ \hline
M$_{\text{r}}$   & 214.22 \\ \hline
Crystal system, space group & tetragonal, P4/mbm \\ \hline
Pressure & 175(10)\,GPa \\ \hline
a,c\,(\AA) & 4.0013(12), 2.442(2) \\ \hline
Volume & 39.10(4) \\ \hline
Z& 2\\ \hline
Radiation type & Synchrotron, $\lambda$=0.3092\,\AA \\ \hline
$\mu$ (mm$^{-1}$) & 173.06 \\ \hline
Crystal size (mm$\times$mm$\times$mm)& 0.001$\times$0.001$\times$0.001 \\ \hline
\multicolumn{2}{l}{\bf Data collection}\\
Diffractometer & ID11 \@ ESRF \\ \hline
Absorption correction & 
\begin{tabular}{l}
Multi-scan CrysAlis PRO 1.171.41.116a (Rigaku Oxford Diffraction, 2021)\\
Empirical absorption correction using spherical harmonics \\ 
implemented in SCALE3 ABSPACK scaling algorithm.
\end{tabular} \\ \hline
T$_{\text{min}}$,T$_{\text{max}}$ & 0.3165, 1.000 \\ \hline
\begin{tabular}{l}
No. of measured, independent and\\
observed [I$>$2$\sigma$(I)] reflections
\end{tabular} & 167, 59, 37 \\ \hline
R$_{\text{int}}$ & 0.060 \\ \hline
$(\sin{\Theta}/\lambda)_{\text{max}}$ (\AA$^{-1}$) & 0.936 \\ \hline
\multicolumn{2}{l}{\bf Refinement}\\
R[F$^2>2\sigma$(F$^2$)], wR(F$^2$), S & 0.043, 0.107, 1.02 \\ \hline
No. of reflections & 59 \\ \hline
No. of parameters & 6 \\ \hline
\multicolumn{2}{l}{\bf Crystal structure}\\
&Re, N\\ \hline
Wyckoff site & 2b, 4g \\ \hline
Coordinates & (0, 0, 0.5), (0.618(9), 0.118(9), 0) \\
\end{tabular}
\caption{Data collection, refinement and crystal structure data for the P4/mbm polymorph of ReN$_2$}
\label{tab_3}
\end{table}

\begin{figure}
\includegraphics[width=8cm]{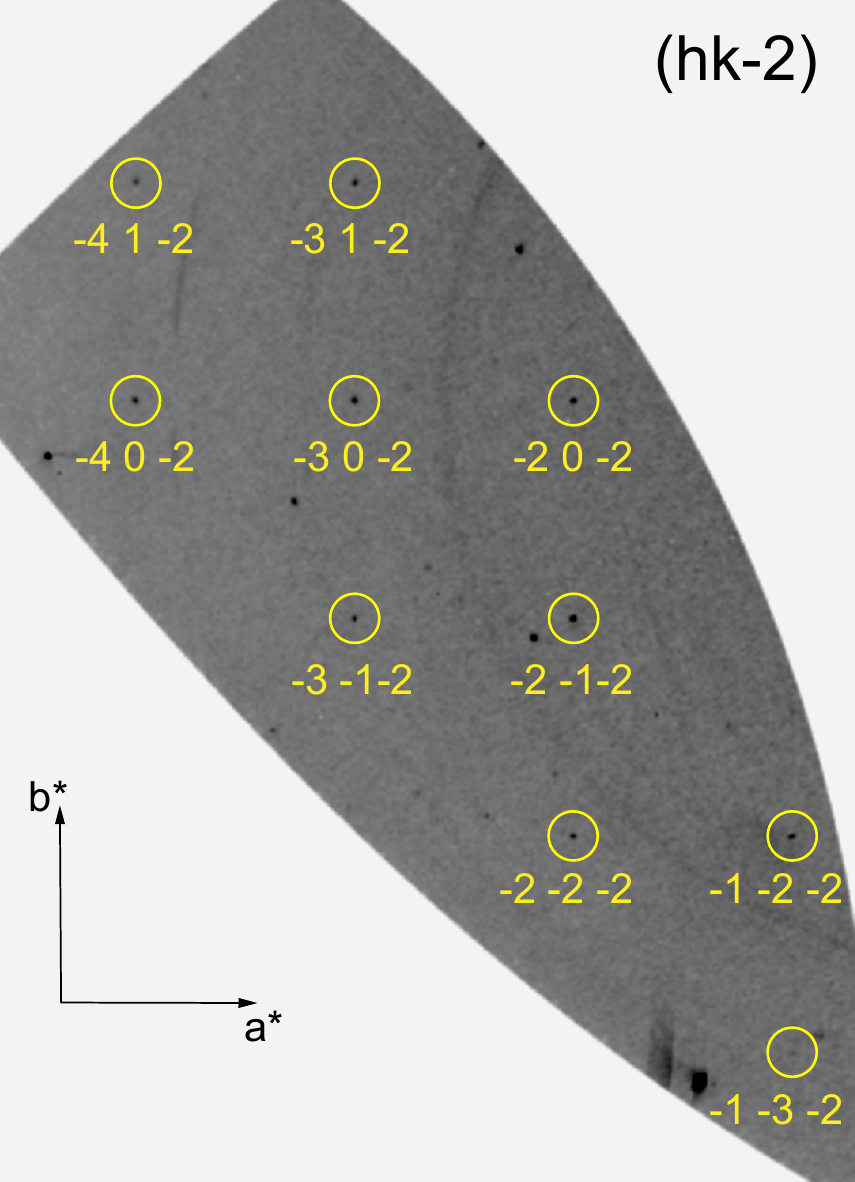}
\caption{
Reconstructed precession images showing the (hk-2) reciprocal lattice plane of P4/mbm-ReN$_2$ at $\approx$175 GPa. The indexed reflections are
encircled. Non-indexed reflections correspond to other grains of ReN$_2$, which are present in the sample.}
\label{fig_exp}
\end{figure}

The characterization of the high pressure synthesized sample has shown that at $\approx$175\,GPa
ReN$_2$ indeed crystallizes in the tetragonal space group (P4/mbm, No. 127) with Re and N occupying Wyckoff sites 2b and 4g, respectively.
See Table \ref{tab_3} and supplementary CIF file \cite{supplementary} for details.
Figure \ref{fig_exp} shows the X-ray reflections corresponding to the (hk-2) reciprocal lattice plane of the P4/mbm phase.
Re atoms are coordinated by eight nitrogen atoms forming ReN$_8$ rectangular prisms. These prisms are
stacked along the $c$-axis (short edge of the rectangular prism) sharing faces and forming infinite columns. The columns also share
common edges and additionally interconnected via N-N bonds as shown in Fig.\ref{fig_1}d). The refined N-N distance of 1.34 \AA
is close to the expected value of a single N-N bond at this pressure \cite{eremets_single-bonded_2004} suggesting that nitrogen
forms a pernitride anion [N-N]$^{4-}$, while Re has an oxidation state +4.

To understand the electronic properties of ReN$_2$ polymorphs at 0 GPa we have calculated
the electronic structure of all three phases. The calculated total and partial density of states
(DOS) are shown in Fig.\ref{fig_3} a), b) and c). For each phase the first panel shows the total DOS curves including the
total rhenium 5d partial DOS as shaded curves. One should note here that the tetragonal polymorph has 
half as many atoms than the other two, therefore the twice of DOS is plotted.
The second panel compares the nitrogen 2p partial DOS for the differnt types (N1 and N2)
nitrogens. In comparison, the total DOS
of the three structures show different characteristics at the Fermi level. One observes nearly semimetallic behavior
for P2$_1$/c and significantly more typical metalic behaviour for the two competing phases. For
P2$_1$/c phase the calculations predict low DOS value at Fermi energy. In the case of C2/m, one sees instead a peak
close to the Fermi energy, though the Fermi level is located in a valley between the two peaks. This could
indicate smaller contribution from the one-electron energy to the structural stability of the C2/m phase relative to
P2$_1$/c phase. In comparison, for P4/mbm one observes a finite value of the total DOS with a plateau in a vicinity
of the Fermi energy. It also indicates smaller contribution from the one-electron term to the structural stability
of the phase. For each of the polymorphs, one notes significant hybridization between Re 5d and N 2p orbitals.
However, the strongly distorted trigonal prismatic local environments
of the Re atoms do not allow any deeper analysis of the Re 5d orbitals using crystal field theory.

\begin{figure}
\includegraphics[width=8cm]{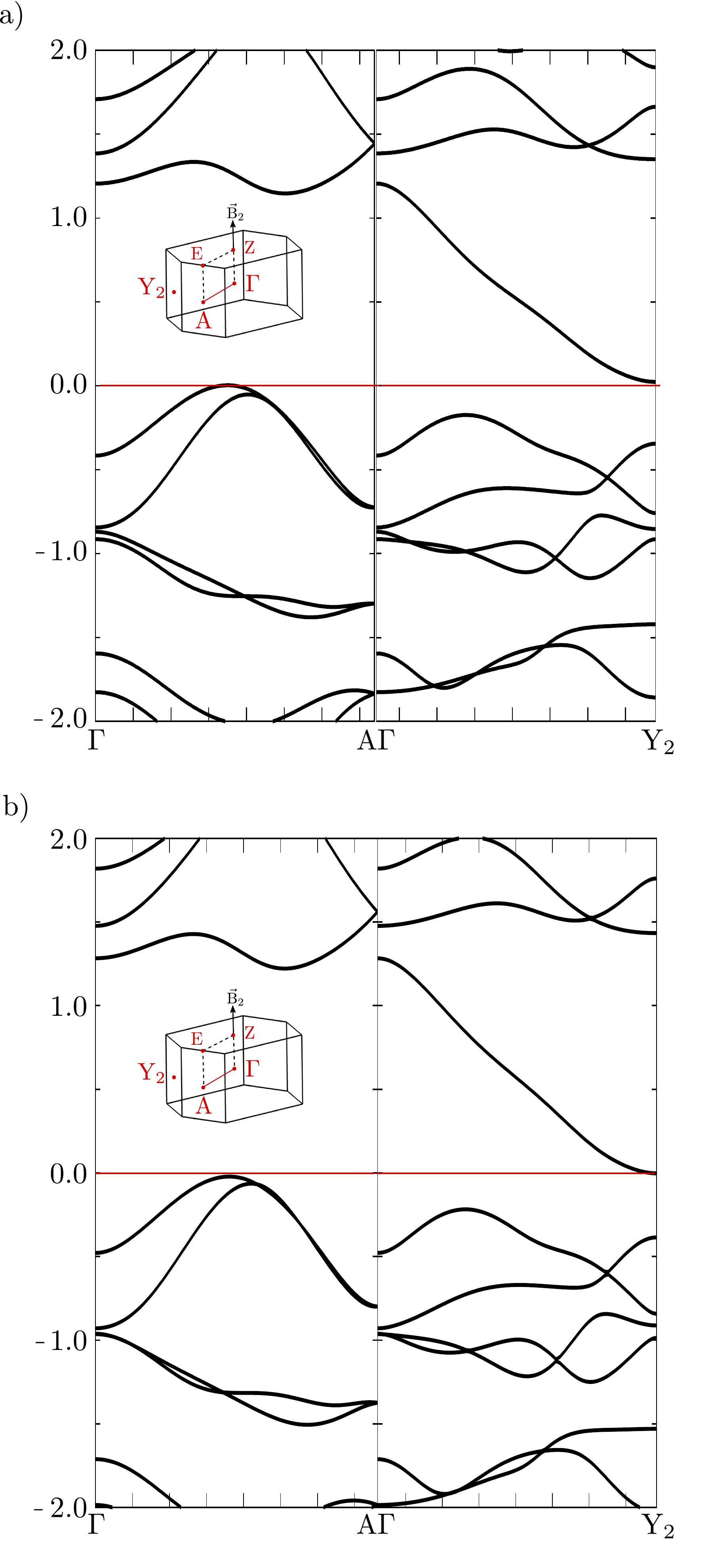}
\caption{Electronic band structure of P2$_1$/c ReN$_2$ at p=0 GPa a), and 23\,GPa b).
The lines shows the electronic bands parallel to the $\Gamma$-A and $\Gamma$-Y$_2$
paths in the Brillouin zone.}
\label{fig_4}
\end{figure}

To analyze the nearly semimetallic DOS of the P2$_1$/c phase we have calculated the electronic
band structure in the Brillouin zone parallel to the $\Gamma$-A and $\Gamma$-Y$_2$ paths, 
see the inset figure of the Brillouin zone in Fig.\,\ref{fig_4}. Fig.\,\ref{fig_4}a) shows the band
structures at ambient pressure while b) shows it for 23\,GPa. One observes that the unoccupied band at Y$_2$
and the occupied band along $\Gamma$-A line could be responsible for the semimetallic behavior
at p=0\,GPa. However, there are small electron pockets along the lines in the $\Gamma$-A-E-Z plane
(not shown) which are responsible for the finite DOS at the Fermi energy. Interestingly, with
increasing pressure (Fig.\,\ref{fig_4}b) the band at Y$_2$ passes the Fermi energy. Besides this, one sees
that the band along the  $\Gamma$-A path crosses the Fermi energy, but not in the $\Gamma$-A-Y$_2$ plane, where
the valence band just touches the Fermi energy. Accrodingly, one expects that the Fermi surface consists of
a simple hole pocket around the $\Gamma$-A line at low pressures, while at higher pressure an additional electron pocket
is expected around the $Y_2$ point in the Brillouin zone. Thus, an electronic topological transition (ETT),
that is the change of the Fermi surface topology \cite{ETT} should be observed with increased pressure.

\begin{figure}
\includegraphics[width=9.5cm]{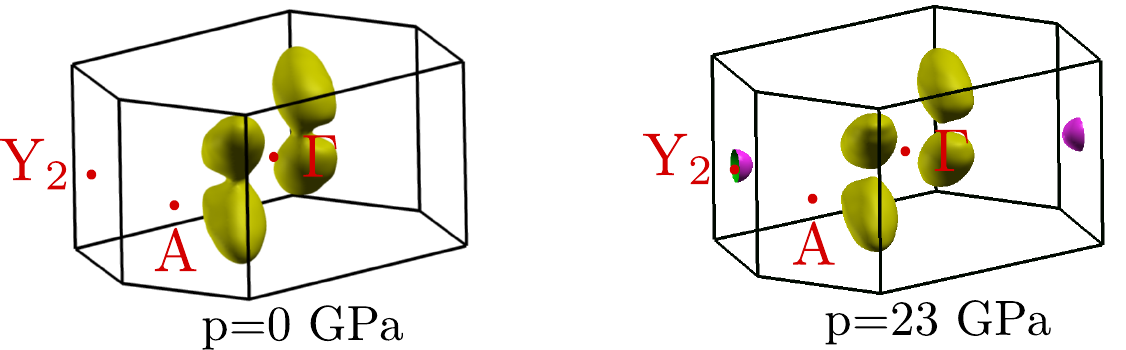}
\caption{Fermi surfaces of the P2$_1$/c phase of ReN$_2$ at 0 and 23\,GPa.}
\label{fig_5}
\end{figure}

To investigate the topology of the Fermi surface and to calculate the pressure at which ETT occurs, we have increased
the accuracy of the calculations by increasing the density of k points in the electronic structure calculations.
We have selected the k point sampling $(26\times14\times20)$
which provides sufficient resolution for the study of the ETT and its influence on the materials properties.
Figure \ref{fig_5} shows the Fermi surface of the P2$_1$/c phase of ReN$_2$ at 0 GPa
and 23 GPa. The figure underlines the appearence of the additional electron pocket around $Y_2$ with increasing pressure.
Importantly, one sees the second ETT associated with a disruption of the neck between the two shits of the Fermi surface
slightly off from the $\Gamma$-A line.
The ETT is connected to the band which touches the Fermi energy along this line at p=0 GPa (Fig.\,\ref{fig_5}) and shifts below
it with increasing pressure (Fig.\,\ref{fig_5}b). Based on the
chosen 1 GPa pressure grid the calculations have shown that the two ETTs occur at 18$\pm0.5$ GPa.

\begin{figure}
\includegraphics[width=9.5cm]{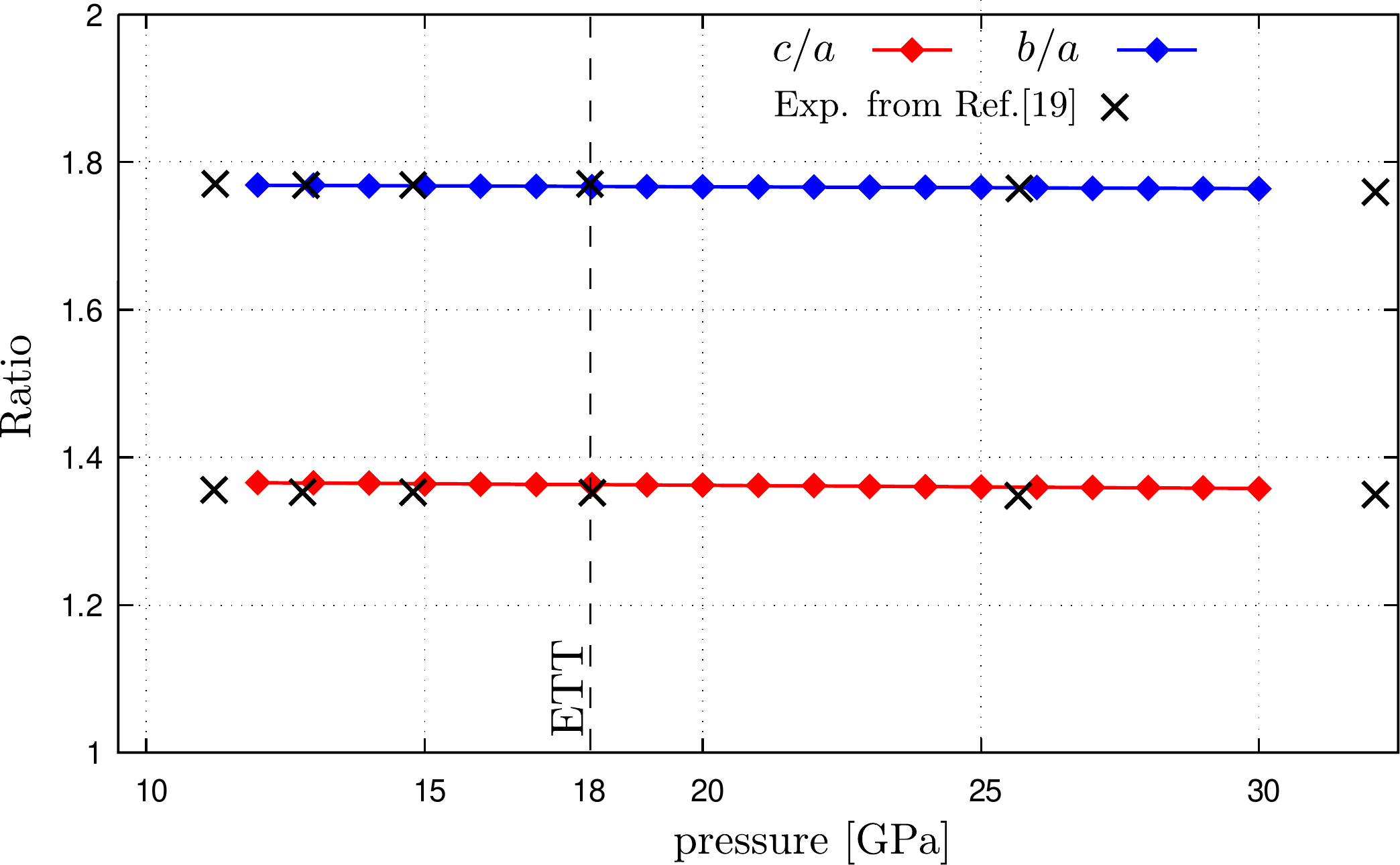}
\caption{Relative lattice parameters of the P2$_1$/c phase of ReN$_2$ calculated with a k-mesh of $(26\times14\times20)$.}
\label{fig_6}
\end{figure}

Experimental identification of the pressure-induced ETT is a non-trivial task, as exemplified by cases of Zn
\cite{kenichi_structural_1997,kenichi_absence_1999,klotz_is_1998,steinle-neumann_absence_2001}
Os
\cite{kenichi_bulk_2004,occelli_experimental_2004,armentrout_incompressibility_2010,godwal_high-pressure_2012,dubrovinsky_most_2015}
and Fe
\cite{glazyrin_importance_2013}
The point is that the thermodynamic potential and its first derivatives are not affected by an ETT, the
second derivatives may show weak square-root shaped peculiarities, while the strong peculiarities are
observed only for the third derivatives of the thermodynamic potential, leading to a classification of ETTs 
as the so-called "2 1/2" order phase transitions. Indeed, Fig.\,\ref{fig_6} shows that, as expected,
the pressure dependence of the lattice parameters ratios c/a and b/a obtained in highly converged calculations
at T=0\,K does not show any peculiarities. However, as pointed out in \cite{glazyrin_importance_2013}, the ETT should
lead to peculiarities of the thermal expansion, and it can show up in the lattice parameters ratios measured
at finite temperatures due to anisotropy of the thermal expansion. The effect was indeed observed experimentally
in hcp Fe \cite{glazyrin_importance_2013} and Os \cite{dubrovinsky_most_2015}. Interestingly, comparing the calculated zero
temperature lattice parameters ratios in Fig.\,\ref{fig_6}  with room temperature
experiment of Ref.\cite{bykov_high-pressure_2019} we observe good agreement between
the two data sets. But the experimental information at pressure around 18 GPa is, unfortunately, missing.
Therefore, careful examination of the lattice parameters of P2$_1$/c phase of ReN$_2$ can be used to investigate
the effect of the predicted ETT on the properties of this compound. 

\section{Conclusions}
We have investigated the thermodynamic and electronic properties of the novel P2$_1$/c phase of ReN$_2$
in comparison with previously suggested, competing phases. 
Our density functional theory calculations at T=0\,K have shown
that the P2$_1$/c phase is the most stable polymorph of the three studied modifications of the compound 
up to $\approx$170\,GPa. Above this pressure the tetragonal P4/mbm becomes more stable.
This calculation is supported by the experiment.
Using the quasi-harmonic approximation we have shown that the P2$_1$/c phase is aslo stable phase up to
1000\,K at p=100 GPa. Moreover, our electronic structure calculations have
shown that two nearly co-existing electronic topological transitions occur in the P2$_1$/c phase of ReN$_2$
with increasing pressure. We propose additional experiments that should verify the theoretically predicted ETT.
\section{Acknowledgment}
We are grateful to Prof. M. I. Katsnelson for useful discussions.
Support from the Knut and Alice Wallenberg Foundation (Wallenberg Scholar Grant No. KAW-2018.0194),
the Swedish Government Strategic Research Areas in Materials Science on Functional Materials at
Link\"oping University (Faculty Grant SFO-Mat-LiU No. 2009 00971) and SeRC, the Swedish Research
Council (VR) grant No. 2019-05600 and the VINN Excellence Center Functional Nanoscale Materials
(FunMat-2) Grant 2016–05156 is gratefully acknowledged. 
Electronic structure calculations
were supported by the Russian Science Foundation (Project No. 18-12-00492).
The computations were enabled by resources
provided by the Swedish National Infrastructure for Computing (SNIC) partially funded by the
Swedish Research Council through grant agreement no. 2016-07213.
The experiments were performed on beamline ID11 at the European Synchrotron Radiation Facility (ESRF),
Grenoble, France. We are grateful to Pavel Sedmak at the ESRF for providing assistance in using beamline. 
%
\bibliographystyle{apsrev4-1}
\bibliography{ReN2_elstruct.bib}
\end{document}